\documentclass[aps,prl, reprint, nofootinbib]{revtex4-1}

\pdfoutput=1

\usepackage[bookmarks=false,hyperfootnotes=false]{hyperref}
\usepackage{amsmath}
\usepackage{color}
\usepackage{graphicx}
\usepackage{tikz}

\def\be{\begin{equation}}
\def\ee{\end{equation}}

\newcommand{\nbar}{\bar{n}}

\newcommand{\zcut}{z_{\text{cut}}}
\newcommand{\glob}{\text{G}}

\begin{document}
\title{Precision physics with pile-up insensitive observables}

\author{Christopher Frye}
 	\email{frye@physics.harvard.edu}
	\affiliation{Center for the Fundamental Laws of Nature, Harvard University, Cambridge, MA 02138, USA}

\author{Andrew J.~Larkoski}
 	\email{larkoski@physics.harvard.edu}
	\affiliation{Center for the Fundamental Laws of Nature, Harvard University, Cambridge, MA 02138, USA}
 
\author{Matthew D.~Schwartz}
 	\email{schwartz@physics.harvard.edu}
	\affiliation{Center for the Fundamental Laws of Nature, Harvard University, Cambridge, MA 02138, USA}

\author{Kai Yan}
 	\email{kyan@physics.harvard.edu}
	\affiliation{Center for the Fundamental Laws of Nature, Harvard University, Cambridge, MA 02138, USA}

\date{\today}

\begin{abstract}

To deepen the search for beyond the Standard Model physics, the Large Hadron Collider is pushing to higher and higher luminosity. 
At high luminosity, precision physics becomes increasingly difficult due to contamination from additional proton collisions per bunch crossing called pile-up.
In recent years, many methods have been developed to cull this excess mostly low-energy radiation away from important signal regions, but it has
been unclear if these methods were amenable to systematically-improvable theoretical understanding. In this paper, it is shown that one such method,
 soft drop jet grooming, has excellent theoretical properties: it is ultra-local, depending on only radiation within a jet, and it is free
of non-global logarithms. Calculations of the soft drop jet mass and related observables are presented at next-to-next-to-leading logarithmic accuracy matched to next-to-next-to-leading fixed-order in perturbative Quantum Chromodynamics. Once measured at the Large Hadron Collider, precision comparisons between theory and data can be made, essentially independent of the amount of pile-up contamination.

\end{abstract}

\maketitle

\begin{figure}[t]
\centering
\includegraphics[width=0.9\linewidth]{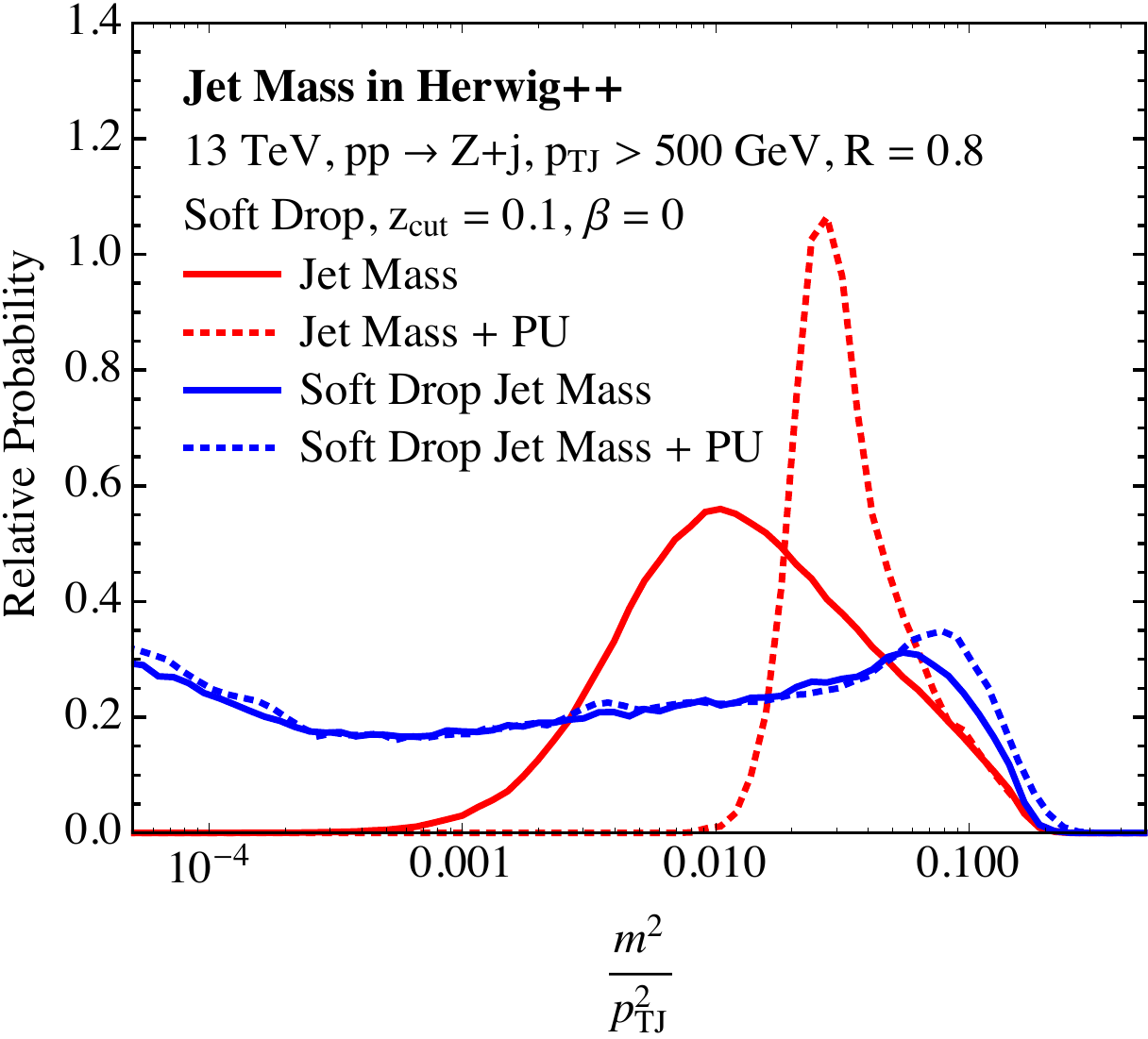}
\caption{
Comparison of the distribution of the jet mass to the jet mass after soft drop grooming the jet as simulated in \textsc{Herwig++}.  Solid curves are the jet mass with no pile-up, and dashed curves are the jet mass with the addition of 25 pile-up vertices, as simulated by the addition of 25 minimum bias events in \textsc{Herwig++}.
}
\label{fig:sdpu}
\end{figure}

As the Large Hadron Collider (LHC) at CERN pushes to higher luminosity, precision comparisons between data and theory become more challenging.
Observables which might be computable if only two protons were colliding become smeared, sometimes beyond recognition, by radiation from
the dozens of protons colliding in the same bunch crossing, known as ``pile-up''
 (see Fig.~\ref{fig:sdpu}). A related smearing comes from the interaction of multiple partons within the same proton collision, called ``underlying event''. 

One approach to handle this extra radiation is to model it, tune the model with data, and incorporate it into theory predictions. Another approach is to calculate it from first principles.
In particular, the underlying event, which entails violation of the simplest form of factorization, is growing as an active area of research in Quantum Chromodynamics (QCD) \cite{Manohar:2012jr,Kasemets:2012pr,Rinaldi:2013vpa,Gaunt:2014ska}. A third approach is to remove it altogether.
Techniques such as jet trimming~\cite{Krohn:2009th}, pruning~\cite{Ellis:2009me}, jet area subtraction~\cite{Cacciari:2008gn}, jet cleansing \cite{Krohn:2013lba}, PUPPI \cite{Bertolini:2014bba}, etc., have already proved essential for many new physics searches. 
Indeed, pile-up removal is necessary for new physics bump-hunts, either data-driven or Monte-Carlo based. Such searches typically do not require extreme precision since backgrounds can be modeled by smooth distributions fit in side-band regions.

There are many situations, however, in which we want to be able to compare data directly to theory, such as in the precision measurement of a coupling or mass. Then, we need to know exactly how the 
pile-up removal algorithms treat all forms of radiation produced in the final state. From the point of view of perturbative QCD, these algorithms are extremely complicated: many involve iterative methods, non-local sampling, and infrared-unsafe information such as 
the energy in charged particles, and are therefore not well-suited to precision theory. So the natural question is: is there an algorithm which is both useful and amenable to precision theoretical study? 
Reasonable criteria for an affirmative answer are the explicit calculation of an observable beyond the level of Monte-Carlo event generators and a demonstration that the calculation can be systematically improved.
In this paper, we show that the soft drop grooming algorithm~\cite{Larkoski:2014wba} and modified mass drop groomer (mMDT) \cite{Dasgupta:2013ihk} has these desired properties. We calculate some soft drop groomed observables to next-to-next-to-leading order (NNLO) with next-to-next-to-leading logarithmic (NNLL) resummation
and provide an all-orders factorization theorem within the framework of soft-collinear effective theory \cite{Bauer:2000yr,Bauer:2001ct,Bauer:2001yt,Bauer:2002nz}.

From an experimental point of view, the key to soft drop is the observation that the contamination radiation is almost entirely of low energy (soft), relative to the energy of the radiation we want to keep, say, in a jet. The algorithm is very simple. We start with a jet, defined according to an infrared and collinear safe algorithm, then recluster the constituents of the jet with the Cambridge/Aachen algorithm \cite{Dokshitzer:1997in,Wobisch:1998wt}.  We then step through the reconstructed branching history, and at a given branching with branches $i$ and $j$, if
\begin{equation}\label{eq:sddef}
\frac{\min[E_i,E_j]}{E_i+E_j} < \zcut\left(\frac{\theta_{ij}}{R}\right)^\beta\,,
\end{equation}
then the softer branch is removed, and the procedure continues to the  next branching that remains.  
The algorithm terminates when a hard branching failing Eq.~\ref{eq:sddef} is found.
Here, $E_i$ is the energy of branch $i$, $\theta_{ij}$ is the relative angle between the branches, and $R$ is the original jet algorithm's radius.  
$\zcut$ and $\beta$ are the parameters of the soft drop groomer, where $\zcut \sim 0.1$ and $\beta = 0$ or $1$ are typical choices.  
For $\beta=0$  (corresponding to mMDT) the constraint is based solely on energy, so all soft radiation, including soft-collinear radiation is dropped.
For $\beta>0$, soft radiation is not dropped if it is sufficiently collinear. 
By discarding low energy radiation, soft drop makes observables almost entirely insensitive to pile-up and underlying event (cf. Fig.~\ref{fig:sdpu}).

From a theoretical point of view, the key to soft drop is that it manages to make observables local -- depending on radiation only in the jets, not throughout the event -- without introducing the dangerous
non-global logarithms \cite{Dasgupta:2001sh} that can spoil factorization and plague many jet observables. To understand how this happens, let us first review an observable that does not factorize nicely: jet mass with an energy veto.  The out-of-jet veto scale $\Omega$ and the jet mass $m$ must be small compared to the center-of-mass collision energy $Q$ to ensure factorization of the cross section.
For an event with $n$ jets, the differential cross section $d\sigma$ of the $n$ jet masses 
$\{m_j\}$ 
factorizes into a convolution
 $d \sigma \sim H \times S \otimes J \otimes \cdots \otimes J$
 with $H$ a hard function, $S$ a soft function
and $J$ a jet function. The hard function only depends on the jet energies $\{Q_j\}$ and the angles between the jets. Each jet function depends on one jet mass, and the soft function $S$ depends on $\Omega$ as well as the small light-cone momentum component of radiation within each jet. Because the
soft function depends on multiple kinematic quantities, and hence multiple scales, one cannot easily resum all of its large logarithms with the renormalization group. It has what are called non-global logarithms.
Moreover, one cannot simply ignore $\Omega$, being inclusive over out-of-jet radiation.  Small $\Omega$ is required for the factorized form of the cross section to be valid and ensure agreement with full QCD in the threshold limit.

Now consider measuring jet masses after soft drop grooming the jets. In the limit that the masses are small, the cross section again factorizes into hard, jet, and soft functions. 
Say a jet has energy $Q_j$ and mass $m_j$ after grooming. Consider some particle within the jet of energy $E$ and at an angle $\theta$ to the jet axis. The invariant mass of this particle
and the jet is then $m_j^2 = 4Q_j E \sin^2\frac{\theta}{2}$. Thus, for the groomed jet mass to be small ($m_j \ll Q_j$), the particle can either be soft and wide angle ($E \sim m_j^2/Q_j$, $\theta \sim 1$), collinear ($\theta \ll 1$), 
or both soft and collinear.
However, for any fixed $\zcut$, in the limit
$m_j \ll Q_j$ where the factorization theorem applies, all the soft wide-angle radiation must fail soft drop ($E Q_j\sim m_j^2 \ll \zcut Q_j^2$). It therefore
only contributes a $\zcut$-dependent normalization factor, $S_\glob(\zcut)$.
Thus after soft drop, the only radiation which contributes to the mass is collinear and the factorization theorem reduces to 
$
d \sigma = H(\{Q_j\}) \times  S_\glob(\zcut) \times J(\zcut, m_1) \otimes \cdots \otimes J(\zcut,m_n)
$.
This is a key result. It implies that the soft drop mass distribution in each jet is universal: it only depends on the jet direction and flavor,
not on any other jet in the event, the underlying event, or any other source of soft radiation such as pile-up.

We can go further. The jet function $J(\zcut, m)$ 
depends on both the soft drop scale $\zcut$ and the jet mass. Thus it could have dangerous non-global logarithms.
However, assuming that $\zcut\ll 1$, collinear radiation which is not soft must be insensitive to $\zcut$, so we can pull an inclusive $\zcut$-independent jet function $J(m)$ out of  $J(\zcut, m)$.
The remaining dependence of $J(\zcut,m)$ on $\zcut$ and $m$ is due to radiation that is both soft and collinear to the jet direction $n^\mu$. 
Being collinear, its momentum $q$ satisfies $q^- = q \cdot \nbar \gg q \cdot n = q^+$, where $\nbar^\mu$ is the direction backwards to the jet. 
Thus $q^- \approx 2 q^0 = 2 E$ and $q^+ \approx m^2/Q$. 
In the case of $\beta=0$, in which Eq.~\ref{eq:sddef} is simply an energy requirement, 
Lorentz invariance then implies that the remaining dependence of $J(\zcut,m)$ is only on the combination $m^2 \zcut$. 
For $\beta>0$, a similar argument shows that the soft and collinear radiation depends on the combination $m^2 \zcut^{\frac{1}{1+\beta}}$. 
Therefore the cross section fully factorizes as
\begin{align}\label{genfact}
\frac{d\sigma}{d m_1 \cdots d m_n } &= H^{IJ}(\{Q_j\},\mu)S^{IJ}_G(\zcut,\mu)  \\
&
\hspace{-2cm}
\times S_C\Big(\frac{m_1^2 \zcut^{\frac{1}{1+\beta}}}{  \mu^2}\Big)\otimes J_1\Big(\frac{m_1}{\mu}\Big) \cdots 
 S_C\Big(\frac{m_n^2 \zcut^{\frac{1}{1+\beta}}}{  \mu^2}\Big)\otimes J_n\Big(\frac{m_n}{\mu}\Big)
\nonumber
\end{align}
with $1,\ldots, n$ indexing the jets and $I,J$ the color structures. 
$S_C$ is a collinear-soft function that describes radiation that is both soft and collinear to the jet direction.

The hard function $H^{IJ}$ and the global soft function $S^{IJ}_\glob$ are complicated: they depend on the number of hard directions in the event and the color structures.  Critically, however, 
neither of these two functions depend on the observables $\{m_j\}$. Thus they only contribute to the overall normalization. Because of this, we can compute the product $ H S_\glob$ by matching
to fixed-order QCD. The remaining soft-collinear functions $S_C$ and jet functions $J$ depend only on a single infrared scale.
Thus all of the logarithms of the $\{m_j\}$ can be resummed using the renormalization group and the factorization theorem is free of non-global logarithms.
Moreover, the collinear-soft and jet functions only depend on jet flavor (quark or gluon), and not on any global structure of the event or on a matrix in color space. So one can even sum over contributions to the cross section
with different numbers of jets and still have a precision prediction for the soft drop groomed jet mass.

To resum the cross section, we need the anomalous dimensions of the various objects. The resummation of logarithms of the groomed jet masses 
requires the anomalous dimensions  only of the inclusive jet function $J$, which is known to at least 2-loops \cite{Neubert:2004dd,Becher:2009th}, and of the collinear-soft functions $S_C$. 
Thus, from a practical perspective, we do not need to separate the hard and global soft functions. Instead we
can compute their product to fixed-order  then
evolve the jet and collinear-soft functions up to the
renormalization group scale at which the fixed order results are computed. 
The anomalous dimension of $S_C$  can be computed  directly, but it is simpler
to derive it using renormalization group invariance of Eq.~\eqref{genfact}

For the simplest case of $e^+e^- \to$ dijets events, $H$ and its anomalous dimension are known to 3 loops~\cite{vanNeerven:1985xr,Matsuura:1988sm}. For soft drop with $\beta = 0$, the anomalous dimension of $S_G$ in this case is closely 
related to the 2-loop anomalous dimension of the soft function with a global energy veto extracted in Ref.~\cite{Chien:2015cka} from the calculation in Ref.~\cite{vonManteuffel:2013vja}.  By calculating additional effects associated with the Cambridge/Aachen reclustering
step of the soft drop algorithm, we find the non-cusp part of the anomalous dimension
of the collinear-soft function at 2-loops:
\begin{align}\label{eq:softnoncusp}
\hspace{-0.34cm}
\gamma_{S_C}^{\beta = 0}\!=\!\left(\frac{\alpha_s}{4\pi}\right)^2\!\!C_F(-17.00 C_F
+36.24C_A+14.84n_f T_F
)
\end{align}

\begin{figure}[t]
\centering
\includegraphics[width=0.95\linewidth]{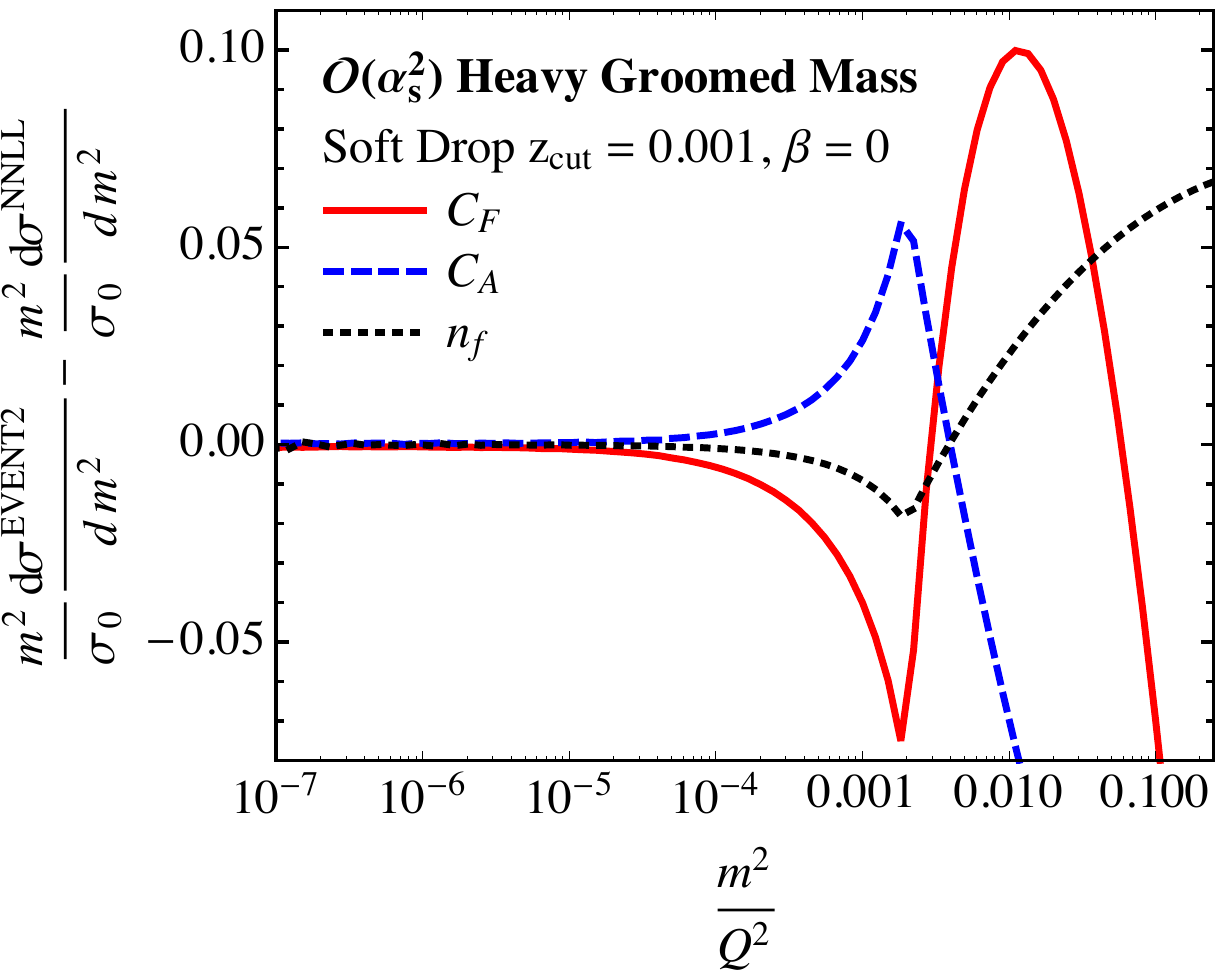}
\caption{
By comparing to the exact 2-loop result, we check that the singular terms in the heavy hemisphere soft drop
groomed mass are reproduced
by the factorization theorem.
}
\label{fig:ev2}
\end{figure}

With this result, we can check the factorization theorem in  Eq.~\eqref{genfact} by expanding to order $\alpha_s^2$. The left-hand side (calculated in full QCD) can be computed
using the program  \textsc{Event2} \cite{Catani:1996vz}. The right-hand side depends on 1-loop results and 2-loop anomalous dimensions. A comparison between the two predictions for the heavy-hemisphere soft drop
groomed mass is shown in Fig.~\ref{fig:ev2}; note the excellent agreement in the singular region.  In this figure, we plot the difference between \textsc{Event2} output and our NNLL result in the three 
different color channels at order $\alpha_s^2$.

For soft drop parameter $\beta \neq 0$, we cannot directly relate the anomalous dimension of the global soft function to calculations in the literature.  Nevertheless, again exploiting renormalization group invariance of Eq.~\ref{genfact}, we can use  \textsc{Event2}  to extract the 2-loop non-cusp anomalous dimension of the collinear-soft function.  Subtracting all known terms at order $\alpha_s^2$ and fitting the $\zcut$ power corrections, for $\beta = 1$, we find
\begin{align}\label{eq:beta1ad}
\gamma_{S_C}^{\beta = 1}&=\left(\frac{\alpha_s}{4\pi}\right)^2C_F[
(-3.0\pm 6)C_F
+(20.5\pm 1)C_A\\
&
\hspace{4cm}
+(8.0\pm 3.5)n_f T_F
]\,.\nonumber
\end{align}
The quoted uncertainties are in part statistical, and in part due to numerical imprecision as \textsc{Event2} probes the deep infrared. 
Our method of extraction is validated by comparing fit values to the exact results for $\beta=0$.

\begin{figure}[t]
\centering
\includegraphics[height=0.8\linewidth]{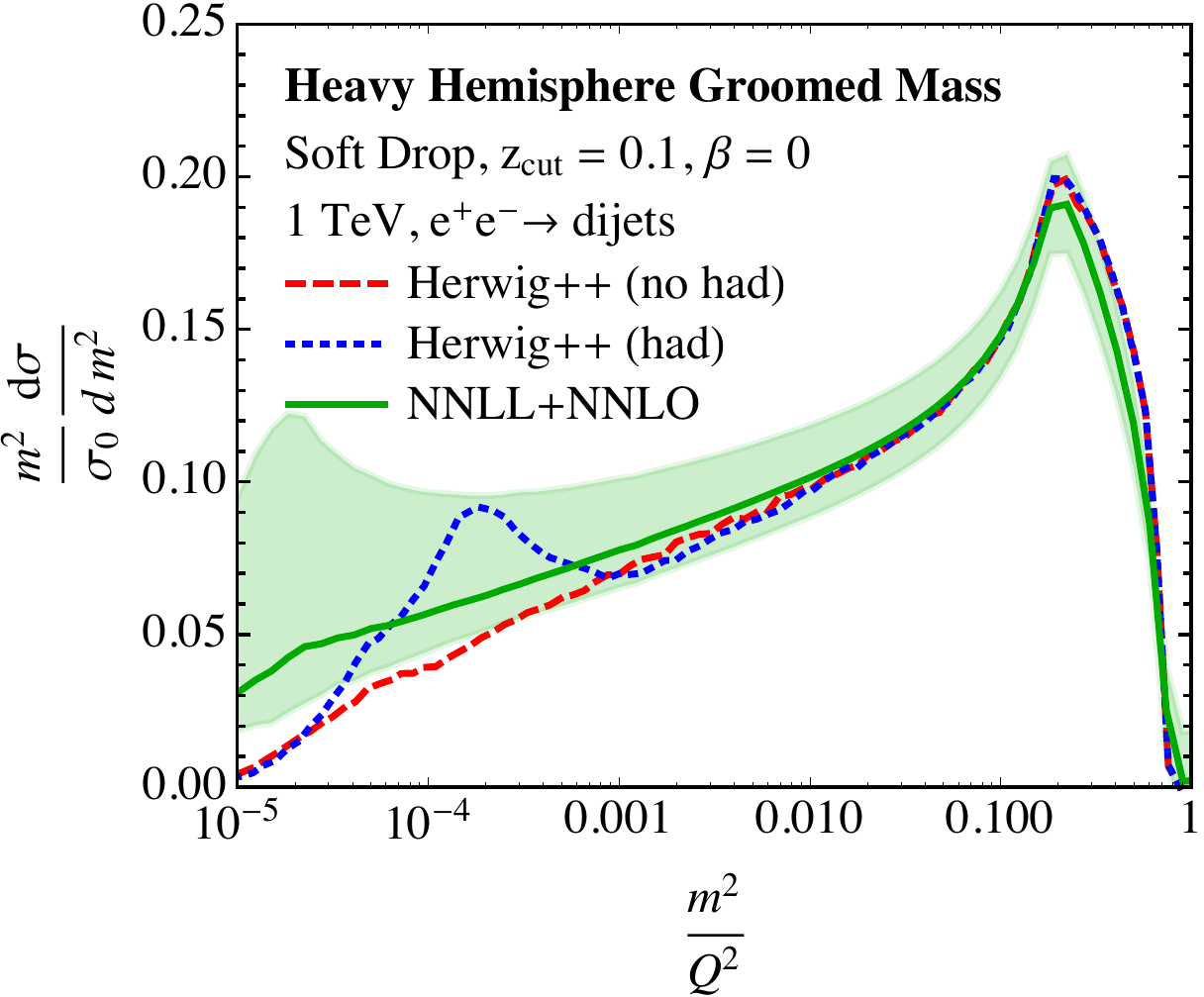}\\
\includegraphics[height=0.8\linewidth]{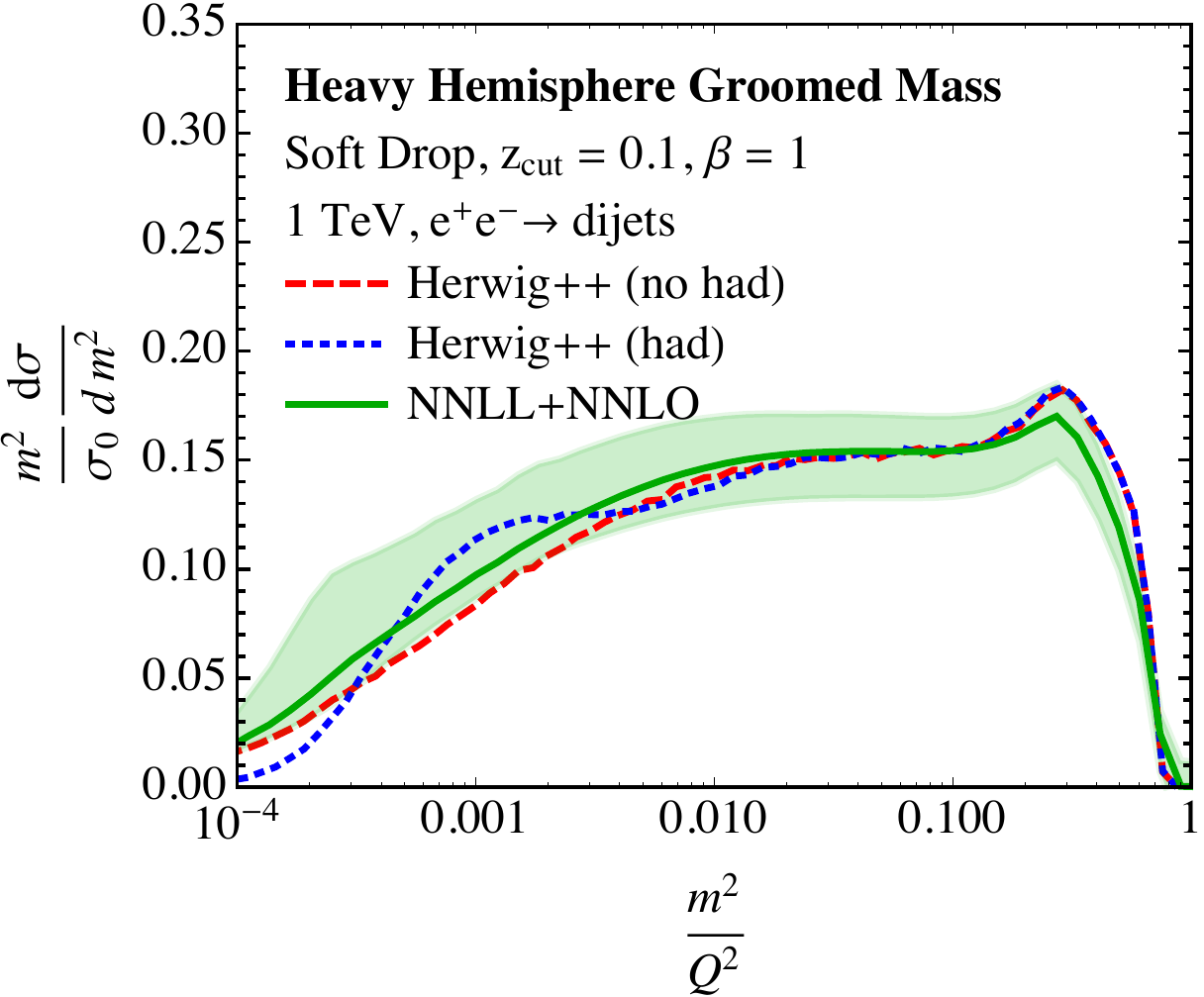}
\caption{
Prediction for the resummed $\beta = 0$ (top) and $\beta = 1$ (bottom) soft drop heavy-hemisphere mass at NNLL matched to NNLO,  and comparison with \textsc{Herwig++}. 
Curves are normalized to have the same total cross section between $0.01 < \frac{m^2}{Q^2} < 1$. 
}
\label{fig:hjmpyth}
\end{figure}

With the complete 2-loop anomalous dimensions, we can compute the soft drop mass distribution to NNLL accuracy. Given also the order $\alpha_s^2$ results from \textsc{Event2}, we
can predict the distribution at NNLL+NNLO order. The resulting calculation is shown in Fig.~\ref{fig:hjmpyth} and compared to \textsc{Herwig++} 2.7.1 \cite{Bahr:2008pv,Bellm:2013hwb} analyzed in \textsc{FastJet} 3.1.3 \cite{Cacciari:2011ma}. The green bands represent estimates of theory uncertainties by varying scales of the functions in the factorization theorem.  For $\beta = 1$, we also include the variation of the 2-loop non-cusp anomalous dimension in Eq.~\ref{eq:beta1ad}.  We see very good agreement. Also note that hadronization has a large effect on the distribution, but only for very small groomed jet masses. This is similar to how hadronization affects the ordinary ungroomed
 heavy jet mass distribution, but pushed to lower scales. Hadronization can be included in the theory prediction using a model shape function \cite{Korchemsky:1999kt,Korchemsky:2000kp}.

Next, let us turn to hadron collisions. The same factorization theorem applies at proton colliders as at $e^+e^-$ colliders,
with the usual modifications of replacing energy $E$ by transverse momentum $p_T$ and polar angle $\theta$ by pseudorapidity $\eta$.
Conveniently, the collinear-soft functions and jet functions are exactly the same, assuming central narrow jets.
Indeed, the distribution of each jet mass is nearly the same as well, up to the normalization factors encoded in the $H S_\glob$ product in Eq.~\eqref{genfact}.
The main substantive change is that at hadron colliders there is a mixture of quark and gluon jets.
At leading power, that is, in the limit that the jets are very jet-like ($m_J \ll p_{TJ}$), the fraction of jets which are quark or gluon is well-defined.
Technically, this can be seen from the lack of mixing between operators with different types of jets (see Ref.~\cite{Gallicchio:2012ez} for some discussion).  
However, one normally cannot compute this fraction in perturbative QCD simply by counting the number of quarks minus the number of antiquarks in the jet. Such a procedure is infrared unsafe:
 adding a soft quark to the jet and a soft anti-quark outside of it would change this counting. However, with soft drop, adding an arbitrarily soft quark to a jet has no effect.
Collinear splittings do not change the number of quarks in a jet either. Thus at leading power for any finite $\zcut$, one {\it can} classify the flavor of a jet in an
infrared safe way by its quark number.

We are therefore led to a simple procedure: compute the relative fractions of quark and gluon soft drop groomed jets at fixed order, and use it to weight the resummed distributions. For example, 
the mass of the highest $p_T$ jet in some event class is given by
\begin{equation}\label{eq:scratch}
 \frac{d\sigma}{d m^2} = \sum_{k=q, g} D_k \, S_{C,k} \otimes J_k   + \frac{d\sigma_{\text{fin.}}}{dm^2}\,.
\end{equation}
Here $D_k$ is the cross section of soft drop groomed quark or gluon jets produced, as extracted from a fixed-order code.  $d\sigma_{\text{fin.}}$ is the finite matching correction,
given by the difference between the fixed-order prediction and the fixed-order expansion of the factorized result.

\begin{figure}[t]
\centering
\includegraphics[height=0.8\linewidth]{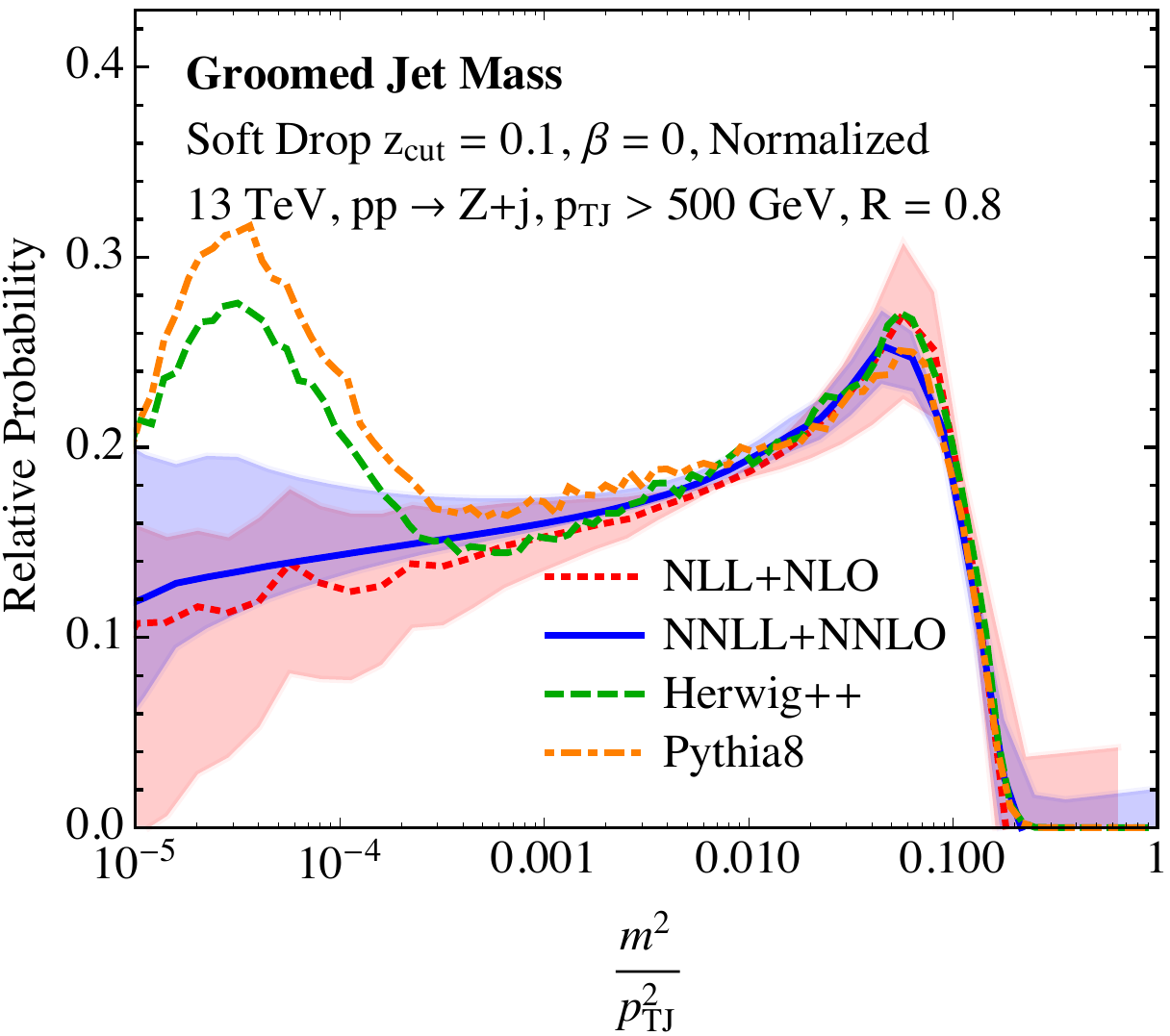}\\
\includegraphics[height=0.8\linewidth]{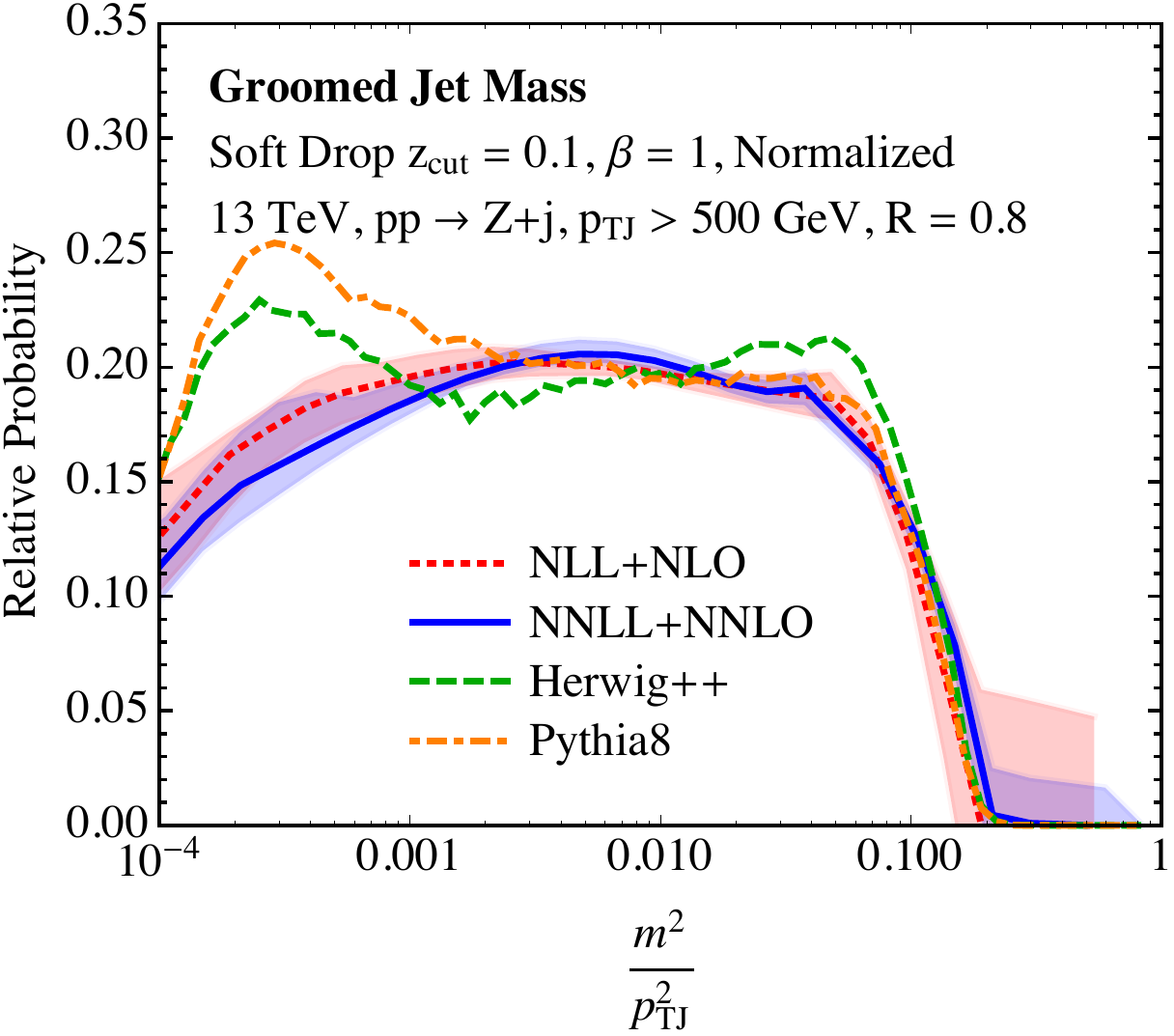}
\caption{
Prediction for the resummed $\beta = 0$ (top) and $\beta =1$ (bottom) soft drop jet mass in $pp\to Z+j$ events at the 13 TeV LHC, as compared to predictions from \textsc{Herwig++} and \textsc{Pythia8},
which include underlying event and hadronization.
The blue and pink shaded bands represent estimates of scale uncertainties and for $\beta = 1$ at NNLL+NNLO also include the uncertainty in the 2-loop anomalous dimension.  
}
\label{fig:ppbeta0}
\end{figure}

As an example application we compute  the soft drop groomed mass of the hardest jet in $pp\to Z+j$ events at the 13 TeV LHC.  We require $p_{TZ} > 300$ GeV, $p_{TJ} > 500$ GeV, $|\eta_Z| < 2.5$ and $|\eta_J| < 2.5$. We find the jets using the anti-$k_T$ algorithm \cite{Cacciari:2008gp} with $R=0.8$, then groom the jets with soft drop. We use \textsc{MCFM} \cite{Campbell:2002tg,Campbell:2003hd} to compute the fixed order distribution and accomplish full NNLO accuracy by demanding that the jets have non-zero mass and using the NLO $pp\to Z+2j$ event generation.  Using \textsc{MCFM}, we can determine the $D_k$ normalization factors and match our NNLL resummed result to NNLO.

The resulting predictions for soft drop groomed jet mass for both $\beta = 0$ and $\beta = 1$ is shown in Fig.~\ref{fig:ppbeta0}.  Because the shape of the distribution is purely determined by collinear emissions in the jet, we have normalized the distributions to integrate to the same value on the range $m^2/p_{TJ}^2 \in[0.001,0.1]$.  We include both NLL+NLO and NNLL+NNLO predictions and their associated uncertainty bands, demonstrating that scale uncertainties decrease in going to higher order.  In these plots, we compare to \textsc{Herwig++} and \textsc{Pythia} 8.210 \cite{Sjostrand:2006za,Sjostrand:2014zea} run with default settings, which includes underlying event and hadronization.
Agreement with our analytic calculation is generally good, though some differences exist.

In conclusion, we have shown how precision jet physics can persist even in the regime of ultra-high luminosity. A particular jet grooming algorithm, soft drop, is not only efficacious, removing essentially
all pile-up, but also lends to remarkable factorization properties. In particular, the factorization formulas for soft drop observables like jet mass are free
of non-global logarithms to all orders. The calculation becomes ultra-local: involving only physics within the jet; the global structure, such as the other jets in the event, only affects the overall normalization. 
This opens the door for detailed comparisons between theory and data even at the ultra-high luminosities planned for the LHC and future colliders.  Calculational details will be presented in a future publication.

We thank Simone Marzani, Ian Moult, Ben Nachman, Duff Neill, Iain Stewart, and Hua-Xing Zhu for discussions.  We thank Andrzej Siodmok for help with generation of events in \textsc{Herwig++}. This research was supported in part by the U.S. Department of Energy, under grant DE-SC0013607.  A.L.~is supported by the U.S. National Science Foundation, under grant PHY--1419008, the LHC Theory Initiative. The computations in this paper were run on the Odyssey cluster supported by the FAS Division of Science, Research Computing Group at Harvard University.

\bibliography{softnnll}

%merlin.mbs apsrev4-1.bst 2010-07-25 4.21a (PWD, AO, DPC) hacked
%Control: key (0)
%Control: author (8) initials jnrlst
%Control: editor formatted (1) identically to author
%Control: production of article title (-1) disabled
%Control: page (0) single
%Control: year (1) truncated
%Control: production of eprint (0) enabled
\begin{thebibliography}{36}%
\makeatletter
\providecommand \@ifxundefined [1]{%
 \@ifx{#1\undefined}
}%
\providecommand \@ifnum [1]{%
 \ifnum #1\expandafter \@firstoftwo
 \else \expandafter \@secondoftwo
 \fi
}%
\providecommand \@ifx [1]{%
 \ifx #1\expandafter \@firstoftwo
 \else \expandafter \@secondoftwo
 \fi
}%
\providecommand \natexlab [1]{#1}%
\providecommand \enquote  [1]{``#1''}%
\providecommand \bibnamefont  [1]{#1}%
\providecommand \bibfnamefont [1]{#1}%
\providecommand \citenamefont [1]{#1}%
\providecommand \href@noop [0]{\@secondoftwo}%
\providecommand \href [0]{\begingroup \@sanitize@url \@href}%
\providecommand \@href[1]{\@@startlink{#1}\@@href}%
\providecommand \@@href[1]{\endgroup#1\@@endlink}%
\providecommand \@sanitize@url [0]{\catcode `\\12\catcode `\$12\catcode
  `\&12\catcode `\#12\catcode `\^12\catcode `\_12\catcode `\%12\relax}%
\providecommand \@@startlink[1]{}%
\providecommand \@@endlink[0]{}%
\providecommand \url  [0]{\begingroup\@sanitize@url \@url }%
\providecommand \@url [1]{\endgroup\@href {#1}{\urlprefix }}%
\providecommand \urlprefix  [0]{URL }%
\providecommand \Eprint [0]{\href }%
\providecommand \doibase [0]{http://dx.doi.org/}%
\providecommand \selectlanguage [0]{\@gobble}%
\providecommand \bibinfo  [0]{\@secondoftwo}%
\providecommand \bibfield  [0]{\@secondoftwo}%
\providecommand \translation [1]{[#1]}%
\providecommand \BibitemOpen [0]{}%
\providecommand \bibitemStop [0]{}%
\providecommand \bibitemNoStop [0]{.\EOS\space}%
\providecommand \EOS [0]{\spacefactor3000\relax}%
\providecommand \BibitemShut  [1]{\csname bibitem#1\endcsname}%
\let\auto@bib@innerbib\@empty
%</preamble>
\bibitem [{\citenamefont {Manohar}\ and\ \citenamefont
  {Waalewijn}(2012)}]{Manohar:2012jr}%
  \BibitemOpen
  \bibfield  {author} {\bibinfo {author} {\bibfnamefont {A.~V.}\ \bibnamefont
  {Manohar}}\ and\ \bibinfo {author} {\bibfnamefont {W.~J.}\ \bibnamefont
  {Waalewijn}},\ }\href {\doibase 10.1103/PhysRevD.85.114009} {\bibfield
  {journal} {\bibinfo  {journal} {Phys. Rev.}\ }\textbf {\bibinfo {volume}
  {D85}},\ \bibinfo {pages} {114009} (\bibinfo {year} {2012})},\ \Eprint
  {http://arxiv.org/abs/1202.3794} {arXiv:1202.3794 [hep-ph]} \BibitemShut
  {NoStop}%
%%CITATION = ARXIV:1202.3794;%%
\bibitem [{\citenamefont {Kasemets}\ and\ \citenamefont
  {Diehl}(2013)}]{Kasemets:2012pr}%
  \BibitemOpen
  \bibfield  {author} {\bibinfo {author} {\bibfnamefont {T.}~\bibnamefont
  {Kasemets}}\ and\ \bibinfo {author} {\bibfnamefont {M.}~\bibnamefont
  {Diehl}},\ }\href {\doibase 10.1007/JHEP01(2013)121} {\bibfield  {journal}
  {\bibinfo  {journal} {JHEP}\ }\textbf {\bibinfo {volume} {01}},\ \bibinfo
  {pages} {121} (\bibinfo {year} {2013})},\ \Eprint
  {http://arxiv.org/abs/1210.5434} {arXiv:1210.5434 [hep-ph]} \BibitemShut
  {NoStop}%
%%CITATION = ARXIV:1210.5434;%%
\bibitem [{\citenamefont {Rinaldi}\ \emph {et~al.}(2013)\citenamefont
  {Rinaldi}, \citenamefont {Scopetta},\ and\ \citenamefont
  {Vento}}]{Rinaldi:2013vpa}%
  \BibitemOpen
  \bibfield  {author} {\bibinfo {author} {\bibfnamefont {M.}~\bibnamefont
  {Rinaldi}}, \bibinfo {author} {\bibfnamefont {S.}~\bibnamefont {Scopetta}}, \
  and\ \bibinfo {author} {\bibfnamefont {V.}~\bibnamefont {Vento}},\ }\href
  {\doibase 10.1103/PhysRevD.87.114021} {\bibfield  {journal} {\bibinfo
  {journal} {Phys. Rev.}\ }\textbf {\bibinfo {volume} {D87}},\ \bibinfo {pages}
  {114021} (\bibinfo {year} {2013})},\ \Eprint {http://arxiv.org/abs/1302.6462}
  {arXiv:1302.6462 [hep-ph]} \BibitemShut {NoStop}%
%%CITATION = ARXIV:1302.6462;%%
\bibitem [{\citenamefont {Gaunt}(2014)}]{Gaunt:2014ska}%
  \BibitemOpen
  \bibfield  {author} {\bibinfo {author} {\bibfnamefont {J.~R.}\ \bibnamefont
  {Gaunt}},\ }\href {\doibase 10.1007/JHEP07(2014)110} {\bibfield  {journal}
  {\bibinfo  {journal} {JHEP}\ }\textbf {\bibinfo {volume} {07}},\ \bibinfo
  {pages} {110} (\bibinfo {year} {2014})},\ \Eprint
  {http://arxiv.org/abs/1405.2080} {arXiv:1405.2080 [hep-ph]} \BibitemShut
  {NoStop}%
%%CITATION = ARXIV:1405.2080;%%
\bibitem [{\citenamefont {Krohn}\ \emph {et~al.}(2010)\citenamefont {Krohn},
  \citenamefont {Thaler},\ and\ \citenamefont {Wang}}]{Krohn:2009th}%
  \BibitemOpen
  \bibfield  {author} {\bibinfo {author} {\bibfnamefont {D.}~\bibnamefont
  {Krohn}}, \bibinfo {author} {\bibfnamefont {J.}~\bibnamefont {Thaler}}, \
  and\ \bibinfo {author} {\bibfnamefont {L.-T.}\ \bibnamefont {Wang}},\ }\href
  {\doibase 10.1007/JHEP02(2010)084} {\bibfield  {journal} {\bibinfo  {journal}
  {JHEP}\ }\textbf {\bibinfo {volume} {1002}},\ \bibinfo {pages} {084}
  (\bibinfo {year} {2010})},\ \Eprint {http://arxiv.org/abs/0912.1342}
  {arXiv:0912.1342 [hep-ph]} \BibitemShut {NoStop}%
%%CITATION = ARXIV:0912.1342;%%
\bibitem [{\citenamefont {Ellis}\ \emph {et~al.}(2010)\citenamefont {Ellis},
  \citenamefont {Vermilion},\ and\ \citenamefont {Walsh}}]{Ellis:2009me}%
  \BibitemOpen
  \bibfield  {author} {\bibinfo {author} {\bibfnamefont {S.~D.}\ \bibnamefont
  {Ellis}}, \bibinfo {author} {\bibfnamefont {C.~K.}\ \bibnamefont
  {Vermilion}}, \ and\ \bibinfo {author} {\bibfnamefont {J.~R.}\ \bibnamefont
  {Walsh}},\ }\href {\doibase 10.1103/PhysRevD.81.094023} {\bibfield  {journal}
  {\bibinfo  {journal} {Phys.Rev.}\ }\textbf {\bibinfo {volume} {D81}},\
  \bibinfo {pages} {094023} (\bibinfo {year} {2010})},\ \Eprint
  {http://arxiv.org/abs/0912.0033} {arXiv:0912.0033 [hep-ph]} \BibitemShut
  {NoStop}%
%%CITATION = ARXIV:0912.0033;%%
\bibitem [{\citenamefont {Cacciari}\ \emph
  {et~al.}(2008{\natexlab{a}})\citenamefont {Cacciari}, \citenamefont {Salam},\
  and\ \citenamefont {Soyez}}]{Cacciari:2008gn}%
  \BibitemOpen
  \bibfield  {author} {\bibinfo {author} {\bibfnamefont {M.}~\bibnamefont
  {Cacciari}}, \bibinfo {author} {\bibfnamefont {G.~P.}\ \bibnamefont {Salam}},
  \ and\ \bibinfo {author} {\bibfnamefont {G.}~\bibnamefont {Soyez}},\ }\href
  {\doibase 10.1088/1126-6708/2008/04/005} {\bibfield  {journal} {\bibinfo
  {journal} {JHEP}\ }\textbf {\bibinfo {volume} {04}},\ \bibinfo {pages} {005}
  (\bibinfo {year} {2008}{\natexlab{a}})},\ \Eprint
  {http://arxiv.org/abs/0802.1188} {arXiv:0802.1188 [hep-ph]} \BibitemShut
  {NoStop}%
%%CITATION = ARXIV:0802.1188;%%
\bibitem [{\citenamefont {Krohn}\ \emph {et~al.}(2014)\citenamefont {Krohn},
  \citenamefont {Schwartz}, \citenamefont {Low},\ and\ \citenamefont
  {Wang}}]{Krohn:2013lba}%
  \BibitemOpen
  \bibfield  {author} {\bibinfo {author} {\bibfnamefont {D.}~\bibnamefont
  {Krohn}}, \bibinfo {author} {\bibfnamefont {M.~D.}\ \bibnamefont {Schwartz}},
  \bibinfo {author} {\bibfnamefont {M.}~\bibnamefont {Low}}, \ and\ \bibinfo
  {author} {\bibfnamefont {L.-T.}\ \bibnamefont {Wang}},\ }\href {\doibase
  10.1103/PhysRevD.90.065020} {\bibfield  {journal} {\bibinfo  {journal} {Phys.
  Rev.}\ }\textbf {\bibinfo {volume} {D90}},\ \bibinfo {pages} {065020}
  (\bibinfo {year} {2014})},\ \Eprint {http://arxiv.org/abs/1309.4777}
  {arXiv:1309.4777 [hep-ph]} \BibitemShut {NoStop}%
%%CITATION = ARXIV:1309.4777;%%
\bibitem [{\citenamefont {Bertolini}\ \emph {et~al.}(2014)\citenamefont
  {Bertolini}, \citenamefont {Harris}, \citenamefont {Low},\ and\ \citenamefont
  {Tran}}]{Bertolini:2014bba}%
  \BibitemOpen
  \bibfield  {author} {\bibinfo {author} {\bibfnamefont {D.}~\bibnamefont
  {Bertolini}}, \bibinfo {author} {\bibfnamefont {P.}~\bibnamefont {Harris}},
  \bibinfo {author} {\bibfnamefont {M.}~\bibnamefont {Low}}, \ and\ \bibinfo
  {author} {\bibfnamefont {N.}~\bibnamefont {Tran}},\ }\href {\doibase
  10.1007/JHEP10(2014)059} {\bibfield  {journal} {\bibinfo  {journal} {JHEP}\
  }\textbf {\bibinfo {volume} {10}},\ \bibinfo {pages} {59} (\bibinfo {year}
  {2014})},\ \Eprint {http://arxiv.org/abs/1407.6013} {arXiv:1407.6013
  [hep-ph]} \BibitemShut {NoStop}%
%%CITATION = ARXIV:1407.6013;%%
\bibitem [{\citenamefont {Larkoski}\ \emph {et~al.}(2014)\citenamefont
  {Larkoski}, \citenamefont {Marzani}, \citenamefont {Soyez},\ and\
  \citenamefont {Thaler}}]{Larkoski:2014wba}%
  \BibitemOpen
  \bibfield  {author} {\bibinfo {author} {\bibfnamefont {A.~J.}\ \bibnamefont
  {Larkoski}}, \bibinfo {author} {\bibfnamefont {S.}~\bibnamefont {Marzani}},
  \bibinfo {author} {\bibfnamefont {G.}~\bibnamefont {Soyez}}, \ and\ \bibinfo
  {author} {\bibfnamefont {J.}~\bibnamefont {Thaler}},\ }\href {\doibase
  10.1007/JHEP05(2014)146} {\bibfield  {journal} {\bibinfo  {journal} {JHEP}\
  }\textbf {\bibinfo {volume} {1405}},\ \bibinfo {pages} {146} (\bibinfo {year}
  {2014})},\ \Eprint {http://arxiv.org/abs/1402.2657} {arXiv:1402.2657
  [hep-ph]} \BibitemShut {NoStop}%
%%CITATION = ARXIV:1402.2657;%%
\bibitem [{\citenamefont {Dasgupta}\ \emph {et~al.}(2013)\citenamefont
  {Dasgupta}, \citenamefont {Fregoso}, \citenamefont {Marzani},\ and\
  \citenamefont {Salam}}]{Dasgupta:2013ihk}%
  \BibitemOpen
  \bibfield  {author} {\bibinfo {author} {\bibfnamefont {M.}~\bibnamefont
  {Dasgupta}}, \bibinfo {author} {\bibfnamefont {A.}~\bibnamefont {Fregoso}},
  \bibinfo {author} {\bibfnamefont {S.}~\bibnamefont {Marzani}}, \ and\
  \bibinfo {author} {\bibfnamefont {G.~P.}\ \bibnamefont {Salam}},\ }\href
  {\doibase 10.1007/JHEP09(2013)029} {\bibfield  {journal} {\bibinfo  {journal}
  {JHEP}\ }\textbf {\bibinfo {volume} {1309}},\ \bibinfo {pages} {029}
  (\bibinfo {year} {2013})},\ \Eprint {http://arxiv.org/abs/1307.0007}
  {arXiv:1307.0007 [hep-ph]} \BibitemShut {NoStop}%
%%CITATION = ARXIV:1307.0007;%%
\bibitem [{\citenamefont {Bauer}\ \emph {et~al.}(2001)\citenamefont {Bauer},
  \citenamefont {Fleming}, \citenamefont {Pirjol},\ and\ \citenamefont
  {Stewart}}]{Bauer:2000yr}%
  \BibitemOpen
  \bibfield  {author} {\bibinfo {author} {\bibfnamefont {C.~W.}\ \bibnamefont
  {Bauer}}, \bibinfo {author} {\bibfnamefont {S.}~\bibnamefont {Fleming}},
  \bibinfo {author} {\bibfnamefont {D.}~\bibnamefont {Pirjol}}, \ and\ \bibinfo
  {author} {\bibfnamefont {I.~W.}\ \bibnamefont {Stewart}},\ }\href {\doibase
  10.1103/PhysRevD.63.114020} {\bibfield  {journal} {\bibinfo  {journal}
  {Phys.Rev.}\ }\textbf {\bibinfo {volume} {D63}},\ \bibinfo {pages} {114020}
  (\bibinfo {year} {2001})},\ \Eprint {http://arxiv.org/abs/hep-ph/0011336}
  {arXiv:hep-ph/0011336 [hep-ph]} \BibitemShut {NoStop}%
%%CITATION = HEP-PH/0011336;%%
\bibitem [{\citenamefont {Bauer}\ and\ \citenamefont
  {Stewart}(2001)}]{Bauer:2001ct}%
  \BibitemOpen
  \bibfield  {author} {\bibinfo {author} {\bibfnamefont {C.~W.}\ \bibnamefont
  {Bauer}}\ and\ \bibinfo {author} {\bibfnamefont {I.~W.}\ \bibnamefont
  {Stewart}},\ }\href {\doibase 10.1016/S0370-2693(01)00902-9} {\bibfield
  {journal} {\bibinfo  {journal} {Phys.Lett.}\ }\textbf {\bibinfo {volume}
  {B516}},\ \bibinfo {pages} {134} (\bibinfo {year} {2001})},\ \Eprint
  {http://arxiv.org/abs/hep-ph/0107001} {arXiv:hep-ph/0107001 [hep-ph]}
  \BibitemShut {NoStop}%
%%CITATION = HEP-PH/0107001;%%
\bibitem [{\citenamefont {Bauer}\ \emph
  {et~al.}(2002{\natexlab{a}})\citenamefont {Bauer}, \citenamefont {Pirjol},\
  and\ \citenamefont {Stewart}}]{Bauer:2001yt}%
  \BibitemOpen
  \bibfield  {author} {\bibinfo {author} {\bibfnamefont {C.~W.}\ \bibnamefont
  {Bauer}}, \bibinfo {author} {\bibfnamefont {D.}~\bibnamefont {Pirjol}}, \
  and\ \bibinfo {author} {\bibfnamefont {I.~W.}\ \bibnamefont {Stewart}},\
  }\href {\doibase 10.1103/PhysRevD.65.054022} {\bibfield  {journal} {\bibinfo
  {journal} {Phys.Rev.}\ }\textbf {\bibinfo {volume} {D65}},\ \bibinfo {pages}
  {054022} (\bibinfo {year} {2002}{\natexlab{a}})},\ \Eprint
  {http://arxiv.org/abs/hep-ph/0109045} {arXiv:hep-ph/0109045 [hep-ph]}
  \BibitemShut {NoStop}%
%%CITATION = HEP-PH/0109045;%%
\bibitem [{\citenamefont {Bauer}\ \emph
  {et~al.}(2002{\natexlab{b}})\citenamefont {Bauer}, \citenamefont {Fleming},
  \citenamefont {Pirjol}, \citenamefont {Rothstein},\ and\ \citenamefont
  {Stewart}}]{Bauer:2002nz}%
  \BibitemOpen
  \bibfield  {author} {\bibinfo {author} {\bibfnamefont {C.~W.}\ \bibnamefont
  {Bauer}}, \bibinfo {author} {\bibfnamefont {S.}~\bibnamefont {Fleming}},
  \bibinfo {author} {\bibfnamefont {D.}~\bibnamefont {Pirjol}}, \bibinfo
  {author} {\bibfnamefont {I.~Z.}\ \bibnamefont {Rothstein}}, \ and\ \bibinfo
  {author} {\bibfnamefont {I.~W.}\ \bibnamefont {Stewart}},\ }\href {\doibase
  10.1103/PhysRevD.66.014017} {\bibfield  {journal} {\bibinfo  {journal}
  {Phys.Rev.}\ }\textbf {\bibinfo {volume} {D66}},\ \bibinfo {pages} {014017}
  (\bibinfo {year} {2002}{\natexlab{b}})},\ \Eprint
  {http://arxiv.org/abs/hep-ph/0202088} {arXiv:hep-ph/0202088 [hep-ph]}
  \BibitemShut {NoStop}%
%%CITATION = HEP-PH/0202088;%%
\bibitem [{\citenamefont {Dokshitzer}\ \emph {et~al.}(1997)\citenamefont
  {Dokshitzer}, \citenamefont {Leder}, \citenamefont {Moretti},\ and\
  \citenamefont {Webber}}]{Dokshitzer:1997in}%
  \BibitemOpen
  \bibfield  {author} {\bibinfo {author} {\bibfnamefont {Y.~L.}\ \bibnamefont
  {Dokshitzer}}, \bibinfo {author} {\bibfnamefont {G.~D.}\ \bibnamefont
  {Leder}}, \bibinfo {author} {\bibfnamefont {S.}~\bibnamefont {Moretti}}, \
  and\ \bibinfo {author} {\bibfnamefont {B.~R.}\ \bibnamefont {Webber}},\
  }\href {\doibase 10.1088/1126-6708/1997/08/001} {\bibfield  {journal}
  {\bibinfo  {journal} {JHEP}\ }\textbf {\bibinfo {volume} {08}},\ \bibinfo
  {pages} {001} (\bibinfo {year} {1997})},\ \Eprint
  {http://arxiv.org/abs/hep-ph/9707323} {arXiv:hep-ph/9707323 [hep-ph]}
  \BibitemShut {NoStop}%
%%CITATION = HEP-PH/9707323;%%
\bibitem [{\citenamefont {Wobisch}\ and\ \citenamefont
  {Wengler}(1998)}]{Wobisch:1998wt}%
  \BibitemOpen
  \bibfield  {author} {\bibinfo {author} {\bibfnamefont {M.}~\bibnamefont
  {Wobisch}}\ and\ \bibinfo {author} {\bibfnamefont {T.}~\bibnamefont
  {Wengler}},\ }in\ \href
  {https://inspirehep.net/record/484872/files/arXiv:hep-ph_9907280.pdf} {\emph
  {\bibinfo {booktitle} {{Monte Carlo generators for HERA physics. Proceedings,
  Workshop, Hamburg, Germany, 1998-1999}}}}\ (\bibinfo {year} {1998})\ \Eprint
  {http://arxiv.org/abs/hep-ph/9907280} {arXiv:hep-ph/9907280 [hep-ph]}
  \BibitemShut {NoStop}%
%%CITATION = HEP-PH/9907280;%%
\bibitem [{\citenamefont {Dasgupta}\ and\ \citenamefont
  {Salam}(2001)}]{Dasgupta:2001sh}%
  \BibitemOpen
  \bibfield  {author} {\bibinfo {author} {\bibfnamefont {M.}~\bibnamefont
  {Dasgupta}}\ and\ \bibinfo {author} {\bibfnamefont {G.~P.}\ \bibnamefont
  {Salam}},\ }\href {\doibase 10.1016/S0370-2693(01)00725-0} {\bibfield
  {journal} {\bibinfo  {journal} {Phys. Lett.}\ }\textbf {\bibinfo {volume}
  {B512}},\ \bibinfo {pages} {323} (\bibinfo {year} {2001})},\ \Eprint
  {http://arxiv.org/abs/hep-ph/0104277} {arXiv:hep-ph/0104277 [hep-ph]}
  \BibitemShut {NoStop}%
%%CITATION = HEP-PH/0104277;%%
\bibitem [{\citenamefont {Neubert}(2005)}]{Neubert:2004dd}%
  \BibitemOpen
  \bibfield  {author} {\bibinfo {author} {\bibfnamefont {M.}~\bibnamefont
  {Neubert}},\ }\href {\doibase 10.1140/epjc/s2005-02141-1} {\bibfield
  {journal} {\bibinfo  {journal} {Eur. Phys. J.}\ }\textbf {\bibinfo {volume}
  {C40}},\ \bibinfo {pages} {165} (\bibinfo {year} {2005})},\ \Eprint
  {http://arxiv.org/abs/hep-ph/0408179} {arXiv:hep-ph/0408179 [hep-ph]}
  \BibitemShut {NoStop}%
%%CITATION = HEP-PH/0408179;%%
\bibitem [{\citenamefont {Becher}\ and\ \citenamefont
  {Schwartz}(2010)}]{Becher:2009th}%
  \BibitemOpen
  \bibfield  {author} {\bibinfo {author} {\bibfnamefont {T.}~\bibnamefont
  {Becher}}\ and\ \bibinfo {author} {\bibfnamefont {M.~D.}\ \bibnamefont
  {Schwartz}},\ }\href {\doibase 10.1007/JHEP02(2010)040} {\bibfield  {journal}
  {\bibinfo  {journal} {JHEP}\ }\textbf {\bibinfo {volume} {02}},\ \bibinfo
  {pages} {040} (\bibinfo {year} {2010})},\ \Eprint
  {http://arxiv.org/abs/0911.0681} {arXiv:0911.0681 [hep-ph]} \BibitemShut
  {NoStop}%
%%CITATION = ARXIV:0911.0681;%%
\bibitem [{\citenamefont {van Neerven}(1986)}]{vanNeerven:1985xr}%
  \BibitemOpen
  \bibfield  {author} {\bibinfo {author} {\bibfnamefont {W.~L.}\ \bibnamefont
  {van Neerven}},\ }\href {\doibase 10.1016/0550-3213(86)90165-3} {\bibfield
  {journal} {\bibinfo  {journal} {Nucl. Phys.}\ }\textbf {\bibinfo {volume}
  {B268}},\ \bibinfo {pages} {453} (\bibinfo {year} {1986})}\BibitemShut
  {NoStop}%
%%CITATION = NUPHA,B268,453;%%
\bibitem [{\citenamefont {Matsuura}\ \emph {et~al.}(1989)\citenamefont
  {Matsuura}, \citenamefont {van~der Marck},\ and\ \citenamefont {van
  Neerven}}]{Matsuura:1988sm}%
  \BibitemOpen
  \bibfield  {author} {\bibinfo {author} {\bibfnamefont {T.}~\bibnamefont
  {Matsuura}}, \bibinfo {author} {\bibfnamefont {S.~C.}\ \bibnamefont {van~der
  Marck}}, \ and\ \bibinfo {author} {\bibfnamefont {W.~L.}\ \bibnamefont {van
  Neerven}},\ }\href {\doibase 10.1016/0550-3213(89)90620-2} {\bibfield
  {journal} {\bibinfo  {journal} {Nucl. Phys.}\ }\textbf {\bibinfo {volume}
  {B319}},\ \bibinfo {pages} {570} (\bibinfo {year} {1989})}\BibitemShut
  {NoStop}%
%%CITATION = NUPHA,B319,570;%%
\bibitem [{\citenamefont {Chien}\ \emph {et~al.}(2016)\citenamefont {Chien},
  \citenamefont {Hornig},\ and\ \citenamefont {Lee}}]{Chien:2015cka}%
  \BibitemOpen
  \bibfield  {author} {\bibinfo {author} {\bibfnamefont {Y.-T.}\ \bibnamefont
  {Chien}}, \bibinfo {author} {\bibfnamefont {A.}~\bibnamefont {Hornig}}, \
  and\ \bibinfo {author} {\bibfnamefont {C.}~\bibnamefont {Lee}},\ }\href
  {\doibase 10.1103/PhysRevD.93.014033} {\bibfield  {journal} {\bibinfo
  {journal} {Phys. Rev.}\ }\textbf {\bibinfo {volume} {D93}},\ \bibinfo {pages}
  {014033} (\bibinfo {year} {2016})},\ \Eprint
  {http://arxiv.org/abs/1509.04287} {arXiv:1509.04287 [hep-ph]} \BibitemShut
  {NoStop}%
%%CITATION = ARXIV:1509.04287;%%
\bibitem [{\citenamefont {von Manteuffel}\ \emph {et~al.}(2014)\citenamefont
  {von Manteuffel}, \citenamefont {Schabinger},\ and\ \citenamefont
  {Zhu}}]{vonManteuffel:2013vja}%
  \BibitemOpen
  \bibfield  {author} {\bibinfo {author} {\bibfnamefont {A.}~\bibnamefont {von
  Manteuffel}}, \bibinfo {author} {\bibfnamefont {R.~M.}\ \bibnamefont
  {Schabinger}}, \ and\ \bibinfo {author} {\bibfnamefont {H.~X.}\ \bibnamefont
  {Zhu}},\ }\href {\doibase 10.1007/JHEP03(2014)139} {\bibfield  {journal}
  {\bibinfo  {journal} {JHEP}\ }\textbf {\bibinfo {volume} {03}},\ \bibinfo
  {pages} {139} (\bibinfo {year} {2014})},\ \Eprint
  {http://arxiv.org/abs/1309.3560} {arXiv:1309.3560 [hep-ph]} \BibitemShut
  {NoStop}%
%%CITATION = ARXIV:1309.3560;%%
\bibitem [{\citenamefont {Catani}\ and\ \citenamefont
  {Seymour}(1997)}]{Catani:1996vz}%
  \BibitemOpen
  \bibfield  {author} {\bibinfo {author} {\bibfnamefont {S.}~\bibnamefont
  {Catani}}\ and\ \bibinfo {author} {\bibfnamefont {M.~H.}\ \bibnamefont
  {Seymour}},\ }\href {\doibase 10.1016/S0550-3213(96)00589-5} {\bibfield
  {journal} {\bibinfo  {journal} {Nucl. Phys.}\ }\textbf {\bibinfo {volume}
  {B485}},\ \bibinfo {pages} {291} (\bibinfo {year} {1997})},\ \bibinfo {note}
  {[Erratum: Nucl. Phys.B510,503(1998)]},\ \Eprint
  {http://arxiv.org/abs/hep-ph/9605323} {arXiv:hep-ph/9605323 [hep-ph]}
  \BibitemShut {NoStop}%
%%CITATION = HEP-PH/9605323;%%
\bibitem [{\citenamefont {Bahr}\ \emph {et~al.}(2008)\citenamefont {Bahr},
  \citenamefont {Gieseke}, \citenamefont {Gigg}, \citenamefont {Grellscheid},
  \citenamefont {Hamilton} \emph {et~al.}}]{Bahr:2008pv}%
  \BibitemOpen
  \bibfield  {author} {\bibinfo {author} {\bibfnamefont {M.}~\bibnamefont
  {Bahr}}, \bibinfo {author} {\bibfnamefont {S.}~\bibnamefont {Gieseke}},
  \bibinfo {author} {\bibfnamefont {M.}~\bibnamefont {Gigg}}, \bibinfo {author}
  {\bibfnamefont {D.}~\bibnamefont {Grellscheid}}, \bibinfo {author}
  {\bibfnamefont {K.}~\bibnamefont {Hamilton}},  \emph {et~al.},\ }\href
  {\doibase 10.1140/epjc/s10052-008-0798-9} {\bibfield  {journal} {\bibinfo
  {journal} {Eur.Phys.J.}\ }\textbf {\bibinfo {volume} {C58}},\ \bibinfo
  {pages} {639} (\bibinfo {year} {2008})},\ \Eprint
  {http://arxiv.org/abs/0803.0883} {arXiv:0803.0883 [hep-ph]} \BibitemShut
  {NoStop}%
%%CITATION = ARXIV:0803.0883;%%
\bibitem [{\citenamefont {Bellm}\ \emph {et~al.}(2013)\citenamefont {Bellm}
  \emph {et~al.}}]{Bellm:2013hwb}%
  \BibitemOpen
  \bibfield  {author} {\bibinfo {author} {\bibfnamefont {J.}~\bibnamefont
  {Bellm}} \emph {et~al.},\ }\href@noop {} {\  (\bibinfo {year} {2013})},\
  \Eprint {http://arxiv.org/abs/1310.6877} {arXiv:1310.6877 [hep-ph]}
  \BibitemShut {NoStop}%
%%CITATION = ARXIV:1310.6877;%%
\bibitem [{\citenamefont {Cacciari}\ \emph {et~al.}(2012)\citenamefont
  {Cacciari}, \citenamefont {Salam},\ and\ \citenamefont
  {Soyez}}]{Cacciari:2011ma}%
  \BibitemOpen
  \bibfield  {author} {\bibinfo {author} {\bibfnamefont {M.}~\bibnamefont
  {Cacciari}}, \bibinfo {author} {\bibfnamefont {G.~P.}\ \bibnamefont {Salam}},
  \ and\ \bibinfo {author} {\bibfnamefont {G.}~\bibnamefont {Soyez}},\ }\href
  {\doibase 10.1140/epjc/s10052-012-1896-2} {\bibfield  {journal} {\bibinfo
  {journal} {Eur.Phys.J.}\ }\textbf {\bibinfo {volume} {C72}},\ \bibinfo
  {pages} {1896} (\bibinfo {year} {2012})},\ \Eprint
  {http://arxiv.org/abs/1111.6097} {arXiv:1111.6097 [hep-ph]} \BibitemShut
  {NoStop}%
%%CITATION = ARXIV:1111.6097;%%
\bibitem [{\citenamefont {Korchemsky}\ and\ \citenamefont
  {Sterman}(1999)}]{Korchemsky:1999kt}%
  \BibitemOpen
  \bibfield  {author} {\bibinfo {author} {\bibfnamefont {G.~P.}\ \bibnamefont
  {Korchemsky}}\ and\ \bibinfo {author} {\bibfnamefont {G.~F.}\ \bibnamefont
  {Sterman}},\ }\href {\doibase 10.1016/S0550-3213(99)00308-9} {\bibfield
  {journal} {\bibinfo  {journal} {Nucl. Phys.}\ }\textbf {\bibinfo {volume}
  {B555}},\ \bibinfo {pages} {335} (\bibinfo {year} {1999})},\ \Eprint
  {http://arxiv.org/abs/hep-ph/9902341} {arXiv:hep-ph/9902341 [hep-ph]}
  \BibitemShut {NoStop}%
%%CITATION = HEP-PH/9902341;%%
\bibitem [{\citenamefont {Korchemsky}\ and\ \citenamefont
  {Tafat}(2000)}]{Korchemsky:2000kp}%
  \BibitemOpen
  \bibfield  {author} {\bibinfo {author} {\bibfnamefont {G.~P.}\ \bibnamefont
  {Korchemsky}}\ and\ \bibinfo {author} {\bibfnamefont {S.}~\bibnamefont
  {Tafat}},\ }\href {\doibase 10.1088/1126-6708/2000/10/010} {\bibfield
  {journal} {\bibinfo  {journal} {JHEP}\ }\textbf {\bibinfo {volume} {10}},\
  \bibinfo {pages} {010} (\bibinfo {year} {2000})},\ \Eprint
  {http://arxiv.org/abs/hep-ph/0007005} {arXiv:hep-ph/0007005 [hep-ph]}
  \BibitemShut {NoStop}%
%%CITATION = HEP-PH/0007005;%%
\bibitem [{\citenamefont {Gallicchio}\ and\ \citenamefont
  {Schwartz}(2013)}]{Gallicchio:2012ez}%
  \BibitemOpen
  \bibfield  {author} {\bibinfo {author} {\bibfnamefont {J.}~\bibnamefont
  {Gallicchio}}\ and\ \bibinfo {author} {\bibfnamefont {M.~D.}\ \bibnamefont
  {Schwartz}},\ }\href {\doibase 10.1007/JHEP04(2013)090} {\bibfield  {journal}
  {\bibinfo  {journal} {JHEP}\ }\textbf {\bibinfo {volume} {04}},\ \bibinfo
  {pages} {090} (\bibinfo {year} {2013})},\ \Eprint
  {http://arxiv.org/abs/1211.7038} {arXiv:1211.7038 [hep-ph]} \BibitemShut
  {NoStop}%
%%CITATION = ARXIV:1211.7038;%%
\bibitem [{\citenamefont {Cacciari}\ \emph
  {et~al.}(2008{\natexlab{b}})\citenamefont {Cacciari}, \citenamefont {Salam},\
  and\ \citenamefont {Soyez}}]{Cacciari:2008gp}%
  \BibitemOpen
  \bibfield  {author} {\bibinfo {author} {\bibfnamefont {M.}~\bibnamefont
  {Cacciari}}, \bibinfo {author} {\bibfnamefont {G.~P.}\ \bibnamefont {Salam}},
  \ and\ \bibinfo {author} {\bibfnamefont {G.}~\bibnamefont {Soyez}},\ }\href
  {\doibase 10.1088/1126-6708/2008/04/063} {\bibfield  {journal} {\bibinfo
  {journal} {JHEP}\ }\textbf {\bibinfo {volume} {0804}},\ \bibinfo {pages}
  {063} (\bibinfo {year} {2008}{\natexlab{b}})},\ \Eprint
  {http://arxiv.org/abs/0802.1189} {arXiv:0802.1189 [hep-ph]} \BibitemShut
  {NoStop}%
%%CITATION = ARXIV:0802.1189;%%
\bibitem [{\citenamefont {Campbell}\ and\ \citenamefont
  {Ellis}(2002)}]{Campbell:2002tg}%
  \BibitemOpen
  \bibfield  {author} {\bibinfo {author} {\bibfnamefont {J.~M.}\ \bibnamefont
  {Campbell}}\ and\ \bibinfo {author} {\bibfnamefont {R.~K.}\ \bibnamefont
  {Ellis}},\ }\href {\doibase 10.1103/PhysRevD.65.113007} {\bibfield  {journal}
  {\bibinfo  {journal} {Phys. Rev.}\ }\textbf {\bibinfo {volume} {D65}},\
  \bibinfo {pages} {113007} (\bibinfo {year} {2002})},\ \Eprint
  {http://arxiv.org/abs/hep-ph/0202176} {arXiv:hep-ph/0202176 [hep-ph]}
  \BibitemShut {NoStop}%
%%CITATION = HEP-PH/0202176;%%
\bibitem [{\citenamefont {Campbell}\ \emph {et~al.}(2003)\citenamefont
  {Campbell}, \citenamefont {Ellis},\ and\ \citenamefont
  {Rainwater}}]{Campbell:2003hd}%
  \BibitemOpen
  \bibfield  {author} {\bibinfo {author} {\bibfnamefont {J.~M.}\ \bibnamefont
  {Campbell}}, \bibinfo {author} {\bibfnamefont {R.~K.}\ \bibnamefont {Ellis}},
  \ and\ \bibinfo {author} {\bibfnamefont {D.~L.}\ \bibnamefont {Rainwater}},\
  }\href {\doibase 10.1103/PhysRevD.68.094021} {\bibfield  {journal} {\bibinfo
  {journal} {Phys. Rev.}\ }\textbf {\bibinfo {volume} {D68}},\ \bibinfo {pages}
  {094021} (\bibinfo {year} {2003})},\ \Eprint
  {http://arxiv.org/abs/hep-ph/0308195} {arXiv:hep-ph/0308195 [hep-ph]}
  \BibitemShut {NoStop}%
%%CITATION = HEP-PH/0308195;%%
\bibitem [{\citenamefont {Sjostrand}\ \emph {et~al.}(2006)\citenamefont
  {Sjostrand}, \citenamefont {Mrenna},\ and\ \citenamefont
  {Skands}}]{Sjostrand:2006za}%
  \BibitemOpen
  \bibfield  {author} {\bibinfo {author} {\bibfnamefont {T.}~\bibnamefont
  {Sjostrand}}, \bibinfo {author} {\bibfnamefont {S.}~\bibnamefont {Mrenna}}, \
  and\ \bibinfo {author} {\bibfnamefont {P.~Z.}\ \bibnamefont {Skands}},\
  }\href {\doibase 10.1088/1126-6708/2006/05/026} {\bibfield  {journal}
  {\bibinfo  {journal} {JHEP}\ }\textbf {\bibinfo {volume} {0605}},\ \bibinfo
  {pages} {026} (\bibinfo {year} {2006})},\ \Eprint
  {http://arxiv.org/abs/hep-ph/0603175} {arXiv:hep-ph/0603175 [hep-ph]}
  \BibitemShut {NoStop}%
%%CITATION = HEP-PH/0603175;%%
\bibitem [{Sjo(2015)}]{Sjostrand:2014zea}%
  \BibitemOpen
  \href {\doibase 10.1016/j.cpc.2015.01.024} {\bibfield  {journal} {\bibinfo
  {journal} {Comput. Phys. Commun.}\ }\textbf {\bibinfo {volume} {191}},\
  \bibinfo {pages} {159} (\bibinfo {year} {2015})},\ \Eprint
  {http://arxiv.org/abs/1410.3012} {arXiv:1410.3012 [hep-ph]} \BibitemShut
  {NoStop}%
%%CITATION = ARXIV:1410.3012;%%
\end{thebibliography}%
\end{document}